\documentclass[11pt]{article}
\usepackage{amsmath,amssymb,slashed}

\usepackage{amsfonts}

\newcommand{\hoch}[1]{$\, ^{#1}$}

\makeatletter
\@addtoreset{equation}{section}
\makeatother

\newcommand{\be}{\begin{equation}}
\newcommand{\ee}{\end{equation}}
\newcommand{\bea}{\setlength\arraycolsep{2pt} \begin{eqnarray}}
\newcommand{\eea}{\end{eqnarray}}
\newcommand{\nn}{\nonumber}

\def\bm{\bibitem}
\def\ft#1#2{{\textstyle{\frac{\scriptstyle #1}{\scriptstyle #2} } }}
\def\fft#1#2{{\frac{#1}{#2}}}

\def\0{{\sst{(0)}}}
\def\1{{\sst{(1)}}}
\def\2{{\sst{(2)}}}
\def\3{{\sst{(3)}}}
\def\4{{\sst{(4)}}}
\def\5{{\sst{(5)}}}
\def\6{{\sst{(6)}}}
\def\7{{\sst{(7)}}}
\def\8{{\sst{(8)}}}
\def\sst#1{{\scriptscriptstyle #1}}

\def\ep{{\epsilon}}
\def\del{{\partial}}

\def\Dslash{{\slashed D}}
\def\cDslash{{\slashed {\cal D}}}
\def\im{{\rm i}}

\begin{document}

\begin{flushright}
\hfill{MIFPA-11-25\ \ \ \ KIAS-P11038\ \ \ \ CAS-KITPC/ITP-273}
\end{flushright}

\vspace{5pt}
\begin{center}
{\large {\bf Pseudo-Supergravity Extension of the Bosonic String}}

\vspace{7pt}

H. L\"u\hoch{1,2}, C.N. Pope\hoch{3,4} and Zhao-Long Wang\hoch{5}

\vspace{7pt}

\hoch{1}{\it China Economics and Management Academy\\
Central University of Finance and Economics, Beijing 100081, China}

\vspace{7pt}

\hoch{2}{\it Institute for Advanced Study, Shenzhen
University\\ Nanhai Ave 3688, Shenzhen 518060, China}

\vspace{7pt}

\hoch{3}{\it George P. \& Cynthia Woods Mitchell  Institute
for Fundamental Physics and Astronomy,\\
Texas A\&M University, College Station, TX 77843, USA}

\vspace{7pt}

\hoch{4}{\it DAMTP, Centre for Mathematical Sciences,
 Cambridge University,\\  Wilberforce Road, Cambridge CB3 OWA, UK}

\vspace{7pt}

\hoch{5}{\it School of Physics, Korea Institute for Advanced Study,
Seoul 130-722, Korea}

\vspace{10pt}

\underline{ABSTRACT}
\end{center}

We construct a ``pseudo-supersymmetric" fermionic extension of the
effective action of the bosonic string in arbitrary spacetime
dimension $D$.  The theory is invariant under pseudo-supersymmetry
transformations up to the quadratic fermion order, which is
sufficient in order to be able to derive Killing spinor equations in
bosonic backgrounds, and hence to define BPS type solutions
determined by a system of first-order equations.  The
pseudo-supersymmetric theory can be extended by coupling it to a
Yang-Mills pseudo-supermultiplet.  This also allows us to construct
``$\alpha'$ corrections'' involving quadratic curvature terms.  An
exponential dilaton potential term, associated with the conformal
anomaly for a bosonic string outside its critical dimension, can
also be pseudo-supersymmetrised.

\vspace{15pt}

\thispagestyle{empty}





\newpage

\section{Introduction}

   Supersymmetry confers many nice properties on a theory, including
the existence of BPS solutions that may be easier to construct than
non-supersymmetric ones; positivity properties for the energy;
stability arguments; and so on.  A full calculation of the
invariance of the action to all necessary orders in fermions
(usually, this means to quartic order) is essential in order to
establish the true supersymmetry of the theory.  In practice,
however, for many purposes one does not need to make use of the
complete expressions for the supersymmetry transformation rules in
order to study interesting features of the theory.  For example,
when looking for bosonic solutions that are supersymmetric, it
suffices just to examine the Killing spinor equations that follow by
requiring that the supersymmetry transformations of the fermion
fields vanish in the purely bosonic background.

    Killing spinor equations do not only arise in supersymmetric
theories. For example, one may look for solutions of the Einstein
equations in any dimension $D$ that additionally admit a
covariantly-constant spinor, $D_\mu\eta=0$, or a solution of the
equation $D_\mu\eta =c\Gamma_\mu\eta$. This raises the possibility
that one might construct a ``pseudo-supersymmetric'' extension of
such a bosonic theory, by adding suitable fermionic terms to the
bosonic action.\footnote{Somewhat related ideas, in the context of
scalar/gravity theories, were developed in
\cite{cvlupo,culuva,lupova}, where it was observed that there exist
non-supersymmetrisable theories of gravity coupled to scalars where
nonetheless, the scalar potential could be written in terms of a
``superpotential.''  This was developed further in \cite{freske},
where the concept of ``fake supersymmetry'' for such scalar/gravity
theories was introduced.} The defining property of such an extension
would be that the action should be invariant under
pseudo-supersymmetry transformations of the bosonic and fermionic
fields, provided that one works only to quadratic order in the
fermionic fields.  Obviously one cannot expect the invariance to
work beyond the quadratic order, unless the bosonic theory happens
to be one that can be supersymmetrised in the ordinary way, but
nevertheless the pseudo-supersymmetry can be useful for purposes
such as the construction of special solutions that admit a Killing
spinor.

   A simple concrete example of a pseudo-supersymmetric theory is
Einstein gravity in $D$ dimensions, to which one adds a
``pseudo-gravitino'' field $\psi_\mu$.  Thus one may consider the
Lagrangian
\be
{\cal L} = \sqrt{-g}\,
(R+\ft12 \bar\psi_\mu\Gamma^{\mu\nu\rho}D_\nu\psi_\rho)\,,
\label{einstein}
\ee
for which the action is invariant, modulo terms of quartic order in
fermions, under the pseudo-supersymmetry transformations
\be
\delta\psi_\mu = D_\mu\ep\,,\qquad
\delta e^a_\mu = \ft14 \bar\psi_\mu\Gamma^a\ep\,.\label{einsteinsusy}
\ee
(The spinors should be either Majorana or symplectic-Majorana
according to allowed possibilities in each dimension.  In the
symplectic-Majorana case, the (suppressed) $Sp(2)$ indices are
understood to be contracted with $\ep_{ij}$.) The
pseudo-supersymmetric solutions of this theory are Ricci-flat
metrics admitting covariantly-constant spinors.  It should be
pointed out that the crucial property for the pseudo-supersymmetric
invariance of the Lagrangian (\ref{einstein}) is the following
identity
\begin{equation}
\Gamma^{\mu\nu\rho} [D_\nu, D_\rho] \epsilon = G_{\mu\nu} \Gamma^\nu
\epsilon\,,\label{identity}
\end{equation}
where $G_{\mu\nu}$ is the Einstein tensor.  The vanishing
(\ref{identity}) gives rise to the projected integrability condition
for Killing spinors.

   It was recently proposed that the low-energy effective
action for the bosonic string in $D$ dimensions,
\be
{\cal L}=\sqrt{-g}\, (R-\ft12(\del\phi)^2 -\ft1{12} e^{a\phi}\,
H_{\mu\nu\rho}\,
  H^{\mu\nu\rho})\,,\label{boslag}
\ee
where
\be
a^2 = \fft{8}{D-2}\,,
\ee
might provide another example of a pseudo-supersymmetrisable theory
\cite{lupowa}. The principal motivation for this conjecture was the
observation in \cite{clpgen} that the theory described by
(\ref{boslag}) admits consistent Pauli sphere reductions on $S^3$ or
$S^{D-3}$, in which the lower-dimensional theory comprises a finite
number of fields including all the gauge bosons of $SO(4)$ or
$SO(D-2)$ respectively. Such consistent sphere reductions are very
unusual, and most of the other known examples involve reductions of
supergravity theories, such as the $S^7$ reduction from $D=11$, for
which the existence of Killing spinors in the AdS$_4\times S^7$
background plays a crucial role in the consistency proof in
\cite{dewnic}.  Thus, although there obviously cannot be a true
supergravity explanation for the consistent sphere reductions of the
bosonic string action in a general dimension $D$, it is natural to
conjecture that Killing spinors of a pseudo-supersymmetric extension
of the bosonic theory may play a similar role.  Preliminary results
in \cite{lupowa}, preceded by the consideration of Killing spinors
in theories of gravity with antisymmetric tensors in \cite{luwa0},
and extended to include ${\cal O}(\alpha')$ corrections in
\cite{luwang}, conjectured a system of consistent Killing spinor
equations for which a set of suitable projections of the
integrability conditions implied the bosonic equation of motion of
the bosonic string.

Based on the fact that the low-energy effective action of the
bosonic string admits consistent Killing spinor equations
\cite{lupowa,luwang,llw}, we give an explicit construction of a
pseudo-supergravity extension of the bosonic string action
(\ref{boslag}) in $D$ dimensions. We show that it is indeed
pseudo-supersymmetric at the level of quadratic fermion terms.  We
also show that the theory can be extended further to include a
Maxwell or Yang-Mills pseudo-supermultiplet.  In the special case of
$D=10$ dimensions, our results reduce to the standard ones for
${\cal N}=1$ supergravity coupled to Maxwell or Yang-Mills
\cite{Bvn}, at the level of quadratic fermion terms.  Of course in
the $D=10$ case the theory admits a completion when quartic fermion
terms are added, but in all other dimensions the construction will
break down beyond the quadratic order.

By developing an argument given in \cite{bss,berderoo,berderoo2}, our
results for the Einstein-Yang-Mills pseudo-supergravities can be
extended further to include the addition of order $\alpha'$
corrections proportional to the square of the Riemann tensor.

The paper  is organised as follows.  In section 2, we construct the
pseudo-supersymmetric extension of the bosonic string action
(\ref{boslag}), presenting the explicit form of the action and the
pseudo-supersymmetry transformation rules.  In section 3, we extend
these results further by adding a Maxwell or Yang-Mills gauge
pseudo-supermultiplet. In section 4 we re-express our results in the
string frame, and in section 5 we discuss the
pseudo-supersymmetrisation of curvature-squared corrections at order
$\alpha'$.  In section 6 we make a further addition to the theory,
by considering the conformal anomaly term that appears in th
effective action for the bosonic string in dimensions $D\ne 26$
\cite{cafrmape}.  We show that this also can be
pseudo-supersymmetrised.  The paper ends with conclusions and
discussion in section 7.  In an appendix, we review some basic facts
about Majorana and symplectic-Majorana spinors in general
dimensions, and the symmetries of $\Gamma$-matrix products.

\section{Pseudo-supersymmetric Extension of the Bosonic String}

  Motivated by the form of the Lagrangian for ten-dimensional
${\cal N}=1$ supergravity \cite{Bvn},
we make the following ansatz for the Lagrangian
for a pseudo-supersymmetric generalisation to arbitrary dimension $D$:
\bea
e^{-1} {\cal L}_D &=& R -\ft12 (\del\phi)^2 - \ft1{12} e^{a\phi} H_\3^2 +
  \ft12 \bar\psi_\mu \Gamma^{\mu\nu\rho} D_\nu \psi_\rho +
  \ft12 \bar\lambda\Dslash \lambda   +
e_1 \bar\psi_\mu\Gamma^\nu \Gamma^\mu\lambda \del_\nu\phi \nn\\
&&+
  e^{\fft12 a\phi}[ e_2\bar\psi_\mu\Gamma^{\mu\nu\rho\sigma\lambda}\psi_\lambda
+ e_3 \bar\psi^\nu\Gamma^\rho \psi^\lambda +
   e_4 \bar\psi_\mu \Gamma^{\nu\rho\sigma} \Gamma^\mu\lambda+
e_5 \bar\lambda \Gamma^{\nu\rho\sigma}\lambda] H_{\nu\rho\sigma}
\,,\label{lag}
\eea
where $H_\3=dB_\2$ and the constants $e_1,\ldots,e_5$ will be
determined shortly. All terms except the last occur in the
ten-dimensional supergravity Lagrangian.  The last term tuns out to
be necessary in all dimensions other than ten.  We then make the
ansatz
\bea
\delta\psi_\mu &=& D_\mu\ep + e^{\fft12 a \phi}\, (c_1 \Gamma_{\mu\nu\rho\sigma}
F^{\nu\rho\sigma} + c_2 H_{\mu\nu\rho} \Gamma^{\nu\rho})\ep\,,\nn\\
\delta\lambda &=& c_3\, \Big( (\Gamma^\mu\del_\mu\phi)\,\ep +\fft{a}{12}\,
e^{\fft12 a \phi}\, \Gamma^{\mu\nu\rho} H_{\mu\nu\rho}\, \ep\Big)\,,\nn\\
\delta e^a_\mu &=& \ft14 \bar\psi_\mu \Gamma^a\ep\,,\qquad \hbox{so}\quad
 \delta g_{\mu\nu} = \ft12 \bar\psi_{(\mu}\Gamma_{\nu)}\,\ep\,,\nn\\
\delta \phi &=& c_4\, \bar\ep\lambda\,,\nn\\
\delta B_{\mu\nu} &=& \Big[ c_5\, \bar\ep \Gamma_{[\mu} \psi_{\nu]} +
   c_6\, \bar\ep \Gamma_{\mu\nu}\lambda\Big]\, e^{-\fft12 a \phi}\,,\label{susy}
\eea
for the pseudo-supersymmetry transformation rules, again motivated
by the form of the transformations in the ten-dimensional case.  The
constants $c_1,\ldots,c_6$ will also be determined shortly.  Note
that since pseudo-supersymmetry is only expected to hold up to the
quadratic level in fermions, we do not include any higher-order
fermion terms or torsion in the ans\"atze.

As discussed in the appendix, the fermions are
all assumed to be either Majorana or symplectic-Majorana as appropriate
to the spacetime dimension and the nature of the $\Gamma$-matrix
representation.   It is to be understood that the fermions carry suppressed
$Sp(2)$ doublet indices in the symplectic-Majorana case, contracted with
$\epsilon_{ij}$ in fermion bilinear terms.  The symmetries and spinor types
can be read off from Table 1 in the appendix.

   It is now a straightforward matter to determine all the unknown constants,
by requiring that the action be invariant up to the level of
quadratic fermion terms. Carrying this out, we find
\bea
e_1&=& \fft{\im}{2\sqrt{2\beta}}\,,\quad e_2=-\fft1{48}\,,
  \quad e_3=-\fft18\,,\quad
e_4= -\fft{\im \, a}{24\sqrt{2\beta}}\,,\quad e_5=
-\fft{(D-10)}{48\, (D-2)}\,,\cr 
c_1&=& \fft1{12(D-2)}\,,\quad c_2=-\fft{(D-4)}{8(D-2)}\,,\quad
c_3=\fft{\im\,\sqrt{\beta}}{2\sqrt2}\,,\quad
    c_4=\fft{\im}{2\sqrt{2\beta}}\,,\quad c_5= -\fft12\,,\cr 
c_6 &=& -\fft{\im\, a}{4\sqrt{2\beta}}\,.\label{coeffsd}
\eea
Note that $\beta$, appearing in (\ref{betadef}),
is listed for each dimension and representation
in Table 1 in the appendix.

\section{Addition of a Yang-Mills Pseudo-Supermultiplet}

It was shown in \cite{luwang} that consistent Killing spinor
equations can still be defined when Yang-Mills fields are
introduced.  This implies that we can add a Yang-Mills
pseudo-supermultipet to the Lagrangian (\ref{lag}). We can again
gain insight into the form of the additional terms needed for the
inclusion of a Yang-Mills pseudo-supermultiplet by comparing with
the known ten-dimensional results in \cite{Bvn}. For convenience of
notation, we shall first present the results for the case of an
abelian Maxwell multiplet, and then indicate the necessary
modifications for the non-abelian case.   Thus we are led to propose
that the Lagrangian will be given by
\be
({\cal L}_D \ \hbox{with}\ H_{\mu\nu\rho}\rightarrow \widetilde
H_{\mu\nu\rho})
  + {\cal L}_{\hbox{gauge}}\,,
\ee
where
\be
\widetilde H_{\mu\nu\rho} = H_{\mu\nu\rho} -
              3\alpha_1\,  A_{[\mu}\, \del_\nu A_{\rho]}\,,\label{Fcs}
\ee
and
\bea
e^{-1} {\cal L}_{\rm gauge} &=& \alpha_1\, \Big\{-\ft14 e^{\fft{a}{2}\, \phi}
     F^{\mu\nu}F_{\mu\nu} +
\ft12 \bar\chi \Dslash \chi +
 e_6 \,e^{\fft{a}{4}\, \phi} \, F^{\rho\sigma}\,
 \bar\chi \Gamma^\mu \Gamma_{\rho\sigma}
\psi_\mu \cr 
&&\qquad\quad+ e_7\, e^{\fft{a}{4}\, \phi}\,F_{\mu\nu}\, \bar\chi
\Gamma^{\mu\nu}\lambda + e_8 \, e^{\fft{a}{2}\,\phi}\,
 \widetilde H_{\mu\nu\rho}\, \bar\chi \Gamma^{\mu\nu\rho}\chi\Big\} \,,
\label{gaugelag}
\eea
The transformation rules for the gauge
field $A_\mu$ and the gaugino $\chi$ are taken to be
\be
\delta A_\mu = c_7\, e^{-\fft{a}{4}\, \phi}\, \bar\epsilon\Gamma_\mu\chi
\,,\qquad
\delta \chi = c_8\, e^{\fft{a}{4}\, \phi}\, F_{\mu\nu}\, \Gamma^{\mu\nu}
\, \epsilon\,.
\ee
Additionally, the previous transformation rule for $B_{\mu\nu}$ in (\ref{susy})
is now augmented by
\be
\delta_{\rm extra}\, B_{\mu\nu} = \alpha_1\, c_9\, e^{-\fft{a}{4}\, \phi}\,
\bar\chi \Gamma_{[\mu}\epsilon\, A_{\nu]}\,.
\ee

   Requiring pseudo-supersymmetry of the enlarged system, we now find that
the various constants above are given by
\be c_7=2 c_8=- c_9= \fft1{2\sqrt{2}}\,, \qquad
e_6=-\fft1{4\sqrt{2}}\,,\qquad e_7=\fft{\im\,
a}{8\sqrt{\beta}}\,,\qquad e_8 =-\fft1{48}\,. \ee

  The generalisation to Yang-Mills requires replacing (\ref{Fcs}) by
\be
\widetilde H_{\mu\nu\rho} = H_{\mu\nu\rho} -
            3 \alpha_1\,  {\rm tr}'(A_{[\mu}\, \del_\nu A_{\rho]} -\ft13
  A_{[\mu} A_\nu A_{\rho]})\,,\label{Fcs2}
\ee
taking $F_{\mu\nu}=2\del_{[\mu} A_{\nu]} + [A_\mu,A_\nu]$, taking traces
over all terms bilinear in gauge-algebra valued fields $A_\mu$ and $\chi$,
and Yang-Mills covariantising the usual
Lorentz-covariant derivative $D_\mu\chi=D_\mu(\omega)\chi$ to
\be
{\cal D}_\mu(\omega,A)\chi =D_\mu(\omega)\chi
  + [A_\mu,\chi]\,,
\ee
since $\chi$ is in the adjoint representation of the Yang-Mills
group. Our results for the Maxwell multiplet agree, in ten
dimensions and at the quadratic fermion level, with those in
\cite{Bvn}.  The Yang-Mills generalisation can be found in
\cite{berderoo,berderoo2}. (Partial results for the
non-abelianisation of the Maxwell multiplet can be found in
\cite{chaman}.)

To clarify the notation, note that the Yang-Mills 1-forms are
defined by $A_\1=A^I T_I$, where the generators $T_I$ are
anti-hermitian, obeying the Lie algebra $[T_I, T_J]=f^K{}_{IJ} T_K$,
and normalized in the fundamental representation as ${\rm tr} (T_I
T_J) \equiv \beta \delta_{IJ}$. Then the trace ${\rm tr}'$ is
defined by ${\rm tr}' = \fft{1}{\beta} {\rm tr}$.  Also note that
for simplicity, we have have scaled the Yang-Mills fields so that
its coupling appears to be unity, but with an overall coupling
$\alpha_1$ with the Yang-Mills Lagrangian.

\section{The Lagrangian in the String Frame}

  For some purposes, it is advantageous to make a conformal scaling of the
metric and work in the string frame.  The required transformation of
fields is given by
\bea
e^a_\mu &=& e^{\fft14 a \phi}\, \tilde e^a_\mu\,,\qquad \phi=-a \Phi\,,
\qquad \ep = e^{\fft18 a\phi}\, \tilde\ep\,,\nn\\
\psi_\mu &=& e^{\fft18 a \phi}\,
  (\widetilde\psi_\mu -\fft{\im\, a^2}{8\sqrt\beta}\,
\widetilde\Gamma_\mu\widetilde\lambda)\,,
\qquad \lambda =  \fft{a}{2\sqrt2}\, e^{-\fft18 a \phi}\, \widetilde\lambda\,,
\qquad \chi = e^{-\fft18 a \phi}\, \widetilde\chi\,,\label{stringframe}
\eea
where $\widetilde\Gamma_\mu=\tilde e^a_\mu\, \Gamma_a$.
We find, after dropping the tildes for convenience, that in the string
frame the Lagrangians (\ref{lag}) and (\ref{gaugelag}) become
\bea
e^{-1} {\cal L}_{D, {\rm str}} &=& e^{-2\Phi}\, \Big[ R +
4(\del\Phi)^2 - \ft1{12} \widetilde H^2 +
\ft12\bar\psi_\mu\Gamma^{\mu\nu\rho}D_\nu\psi_\rho -\ft12 \,
\bar\lambda \Dslash\lambda -\im\,\sqrt{\beta}\,
\bar\lambda\Gamma^{\mu\nu}\, D_\mu \psi_\nu
\nn\\
&&\qquad+ \bar\psi_\mu \Gamma^\mu\psi_\rho \del^\rho\Phi +
 \fft{\im}{\sqrt{\beta}}\,
\bar\psi_\mu \Gamma^\nu\Gamma^\mu\lambda \del_\nu\Phi\label{stringlag}\\
&& + \widetilde H_{\nu\rho\sigma} \Big\{-\ft1{48}
\bar\psi_\mu\Gamma^{\mu\nu\rho\sigma\lambda}\, \psi_\lambda - \ft18
\bar\psi^\nu\Gamma^\rho\psi^\sigma + \ft1{48}\,
\bar\lambda\Gamma^{\nu\rho\sigma}\lambda+ \ft{\im}{24\sqrt{\beta}}\,
\bar\psi_\mu \Gamma^{\mu\nu\rho\sigma} \lambda \Big\} \Big]\,,\nn
\eea
and
\bea e^{-1}{\cal L}_{\rm gauge,\, str} &=& \alpha_1\,
e^{-2\Phi}\,{\rm tr}'\Big[
  -\ft14  F^{\mu\nu} F_{\mu\nu} +
\ft12 \bar\chi \cDslash(\omega,A) \chi -
 \ft1{4\sqrt2}\,  F^{\rho\sigma}\,
 \bar\chi \Gamma^\mu \Gamma_{\rho\sigma}
\psi_\mu \nn\\
&&\qquad\qquad+\ft{\im}{4\sqrt{2\beta} }\,F_{\mu\nu}\, \bar\chi
\Gamma^{\mu\nu}\lambda -\ft1{48}\,
 \widetilde H_{\mu\nu\rho}\, \bar\chi \Gamma^{\mu\nu\rho}\chi\Big]\,,
\label{gaugestring}
\eea

   The pseudo-supersymmetry transformation rules in the string frame
turn out to be, after dropping the tildes,
\bea \delta\psi_\mu &=& D_\mu\ep -\ft18 \widetilde H_{\mu\nu\rho}\,
\Gamma^{\nu\rho}\, \ep
 \,,\nn\\
\delta\lambda &=&
 -\im \sqrt{\beta}\Big(\Gamma^\mu \partial_\mu \Phi -
\ft1{12} \Gamma^{\mu\nu\rho} \widetilde H_{\mu\nu\rho}\Big)\epsilon\,,\nn\\
\delta\chi &=& \fft1{4\sqrt{2}}\, F_{\mu\nu}\Gamma^{\mu\nu}\ep\,,\nn\\
\delta e^a_\mu &=& \ft14 \bar\psi_\mu \Gamma^a\ep\,,\nn\\
\delta\Phi &=& -\fft{\im}{8\sqrt{\beta}}\, \bar\ep\, \lambda\,,\nn\\
\delta B_{\mu\nu} &=& -\fft12 \bar\ep \Gamma_{[\mu} \psi_{\nu]} -
\fft{\alpha_1}{2\sqrt{2}}\,\bar\chi \Gamma_{[\mu} \, \ep A_{\nu]}\,,\nn\\
\delta A_\mu &=& \fft1{2\sqrt{2}}\, \bar\ep \Gamma_\mu\chi\,. \eea
Note that $\delta\psi_\mu$ may be
re-expressed in terms of a torsionful connection
as $\delta\psi_\mu = D_\mu(\omega_-)\ep$, where
\be \omega_{\mu\pm}^{ab} \equiv \omega_\mu^{ab} \pm \ft12 \widetilde
H_\mu{}^{ab}\,. \ee

\section{Adding a Riemann$^2$ Term}

It was shown in \cite{berderoo,berderoo2} that having obtained the
ten-dimensional action for ${\cal N}=1$ supergravity coupled to
Yang-Mills, it becomes a straightforward matter to deduce the
structure of ${\cal N}=1$ supergravity with a curvature-squared
correction parameterised by a coefficient $\alpha$. The resulting
action is supersymmetric up to the order $\alpha$.  (See also
\cite{bss}.) We can apply the analogous argument in the case of our
pseudo-supersymmetric extensions of the effective action for the
$D$-dimensional bosonic string. The success is ensured by the
construction of consistent Killing spinor equations with
curvature-squared terms \cite{luwang}. The argument is best
described in the string frame.

The key observation is that the supercovariant gravitino curvature
$\psi^{ab}$, defined from
\be
\psi_{\mu\nu}\equiv D_\mu(\omega_-) \psi_\nu -D_\nu(\omega_-) \psi_\mu\,,
\ee
and the torsionful connection $\omega_{\mu+}^{ab}$ transform in the same
way as an $SO(D-1,1)$ Yang-Mills multiplet $(\chi,A_\mu)$.  Thus we may
make the replacements
\bea
A_\mu &\longrightarrow& \omega_{\mu+}^{ab}\,,\nn\\
F_{\mu\nu} &\longrightarrow& R_{\mu\nu}{}^{ab}(\omega_+)\,,\nn\\
\chi &\longrightarrow& \ft{1}{\sqrt2}\psi^{ab}\,, \eea
in the pseudo-supersymmetric Einstein-Yang-Mills action in the
string frame, and thereby obtain an action for a
pseudo-supersymmetric extension of the $D$-dimensional bosonic
string with a curvature-squared correction. In particular, the
gaugino kinetic term $\ft12 \bar\chi \cDslash(\omega,A)\chi$ in
(\ref{gaugestring}) will now become
\be
\ft12 \bar\psi^{ab} \Gamma^\mu {\cal D}_\mu(\omega,\omega_+)\psi_{ab}\,,
\ee
where
\be
{\cal D}_\mu(\omega,\omega_+)\psi^{ab} =
  D_\mu(\omega)\psi^{ab} +
\omega_{\mu+}^{ac}\, \psi_c{}^b + \omega_{\mu+}^{bc}\, \psi^a{}_c
\,.
\ee
The $R^2$ action is given by
\begin{eqnarray}
e^{-1}{\cal L}_{R^2, {\rm str}} &=& \ft12\alpha_2\,
e^{-2\Phi}\,\Big[
  \ft14  R^{\mu\nu ab}(\omega_+) R_{\mu\nu ab}(\omega_+) -
\ft14 \bar\psi^{ab} \Gamma^\mu {\cal
D}_\mu(\omega,\omega_+)\psi_{ab} \cr &&\qquad\quad + \ft18\,
R^{\rho\sigma ab}(\omega_+)\,
 \bar\psi_{ab} \Gamma^\mu \Gamma_{\rho\sigma}
\psi_\mu -\ft{\im}{8\sqrt{\beta}}\,R_{\mu\nu ab}(\omega_+)\,
\bar\psi^{ab} \Gamma^{\mu\nu}\lambda\cr &&\qquad\quad +\ft1{96}\,
 \widetilde H_{\mu\nu\rho}\, \bar\psi^{ab} \Gamma^{\mu\nu\rho}
 \psi_{ab}\Big]\,.\label{rsquarestring}
\end{eqnarray}
The Bianchi identity for the 3-form field strength is now given by
\begin{equation}
d\widetilde H_\3 = -\ft12\alpha_1 {\rm tr}' (F_\2\wedge F_\2) +
\ft14 \alpha_2 R^{ab}_\2(\omega_+)\wedge R_{ab\2}(\omega_+)\,.
\end{equation}
Note that we have adopted the standard supergravity convention
for the torsional curvature
\begin{equation}
R^{ab}_\2(\omega_\pm) = d\omega_\pm^{ab} + \omega^{a c}_\pm\wedge
\omega_{\pm c}{}^b = \ft12 R_{\mu\nu}{}^{ab}(\omega_\pm)\, dx^\mu
dx^\nu\,.
\end{equation}

It should be pointed out that the pseudo-supersymmetric variation of $\omega_+$ in the
curvature-squared term together with the Chern-Simons modification
of $\widetilde H_\3$ vanishes by the leading order of the equations
of motion, and hence it can be ignored at the $\alpha_2$ order.
This property was shown in \cite{berderoo2} for $D=10$, and in
\cite{andr,luwang} for general dimensions.  The key to prove this is
the Bianchi identity of the torsional curvature, namely
\begin{equation}
\nabla_{[\mu}(\omega,\omega_+)R{}_{\nu\rho]}{}^{ab}(\omega_+)=0\,.
\label{torsionalbianchi1}
\end{equation}
Together with the leading order equations of motion, we find
\begin{equation}
\nabla^{\mu}(\omega_-,\omega_+)\Big[e^{-2\Phi}R_{\mu\nu}{}^{ab}
(\omega_+)\Big]={\cal O}(\alpha_2)\,.\label{torsionalbianchi2}
\end{equation}
Note that the connection in the covariant
derivative is $\omega_+$ for the tangent indices in both above equations;
for spacetime indices, it is $\omega$ in (\ref{torsionalbianchi1}) and
$\omega_-$ in (\ref{torsionalbianchi2}).  These two equations are in
direct parallel to the Bianchi identity and the equation of motion
for the gauge field, namely
\begin{eqnarray}
{\cal D}_{[\rho}(\omega, A) F_{\mu\nu]}=0\,,\qquad {\cal
D}^{\mu}(\omega_-,A)(e^{-2\Phi}F_{\mu\nu})={\cal O}(\alpha_1)\,.
\end{eqnarray}
Thus the verification of the corresponding pesudo-supersymmetric
invariance up to the ${\cal O}(\alpha_2)$ order is analogous to that
of the gauge fields.

The bosonic Lagrangian for the $\alpha'$ corrections was derived in
\cite{luwang} with the requirement that the system admit consistent
Killing spinor equations.  The bosonic Lagrangian is also considered
in \cite{andr} in the context of T-duality.

    To conclude this section, We point out that the
supersymmetry invariance of the ten-dimensional ${\cal N}=1$
supergravity Lagrangian with curvature-squared term is proven in
\cite{berderoo,berderoo2} only up to quadratic order in fermions, as
in our case in general dimensions.

\section{Conformal Anomaly Term}

  If one performs a calculation of the beta functions for the background
fields for a bosonic string in a dimension other than 26, one obtains a
conformal anomaly term for the effective action \cite{cafrmape},
which in the Einstein frame takes the form
\be
 -\ft12 m^2\, e^{-\fft12 a\phi}\sqrt{-g}\,.
\ee
It was shown that the existence of consistent Killing spinor
equations is unspoiled by the conformal anomaly \cite{llw}. We may
now examine whether this term can also be pseudo-supersymmetrised.

   Let us assume for now that we are considering a dimension $D$ for which
$\beta$ is, or may be chosen to be, equal to $-1$
(see Table 1 in the appendix).  We may then consider
adding conformal anomaly terms to the Lagrangian, of the form
\be
e^{-1} {\cal L}_c = -\ft12 m^2 e^{-\fft12 a \phi} +
 e^{-\fft14 a \phi}\, \Big[ e_9\, \bar\psi_\mu \Gamma^{\mu\nu} \psi_\nu
  + e_{10}\, \bar\lambda\lambda + e_{11}\, \bar\psi_\mu\Gamma^\mu\lambda
 + \alpha_1\, e_{12} \bar\chi\chi\Big]
\,.\label{conflag}
\ee
We also
add extra terms in the fermion transformation rules, given by
\be
\delta_{\rm extra} \psi_\mu = c_{10}\, e^{-\fft14 a\phi}\,
 \Gamma_\mu\epsilon\,,\qquad
\delta_{\rm extra} \lambda = c_{11}\, e^{-\fft14 a\phi}\, \epsilon\,.
\label{conftrans}
\ee

   We now find that the action, augmented with the contribution from
(\ref{conflag}), is pseudo-supersymmetric using the augmented transformation
rules, with the coefficients in (\ref{conflag}) and (\ref{conftrans})
determined to be
\bea
e_9 &=& \fft{\im\, m}{4\sqrt2}\,,\qquad
e_{10} = \fft{\im\, m (D-4)}{4\sqrt{2}\, (D-2)}\,,\qquad
e_{11}= \fft{\im\, a m}{8}\,,\qquad e_{12}= \fft{\im\, m}{4\sqrt2}\,,\nn\\
c_{10} &=& -\fft{\im\, m}{2\sqrt2\, (D-2)}\,,\qquad
c_{11}= \fft{\im\, a m}{8}\,.
\eea

  After transforming to the string frame using (\ref{stringframe}), and then
dropping the tildes, we find that the conformal anomaly Lagrangian
(\ref{conflag}) becomes
\be
e^{-1} {\cal L}_{c, {\rm str}} = e^{-2\Phi}\, \Big[
-\ft12 m^2 +\ft{\im}{4\sqrt2}m\Big( \bar\psi_\mu \Gamma^{\mu\nu} \psi_\nu
-2 \bar\psi_\mu\Gamma^\mu\lambda -\bar\lambda \lambda+ \alpha_1\, \bar\chi\chi
  \Big)\Big]
\,.\label{conflag2}
\ee
The transformation rules (\ref{conftrans}) in the string frame are now given by
\begin{equation}
\delta_{\rm extra} \psi_\mu =0\,,\qquad
\delta_{\rm extra} \lambda = \ft{\im\sqrt2}{4} m\, \epsilon\,.
\label{conftrans1}
\end{equation}
Note that in the above derivation, we excluded the Lagrangian for the curvature-squared term (\ref{rsquarestring}).  However, the fact that $\delta \psi_\mu$ is unmodified by the conformal anomaly implies that the curvature-squared and conformal anomaly terms can be pseudo-supersymmetrised simultaneously.  To include (\ref{rsquarestring}), we need to add the following term
\begin{equation}
-\ft{\rm i\,\alpha_2}{8\sqrt2}\, e^{-2\Phi}\, \sqrt{-g}\,\bar\psi^{ab}\psi_{ab}
\end{equation}
to the Lagrangian ${\cal L}_{c, \rm str}$.

  In situations where $\beta=+1$, the terms in (\ref{conflag}) with
coefficients $e_9$ and $e_{10}$ will vanish identically, if the
spinors are Majorana or symplectic-Majorana.  In these cases, one
can still pseudo-supersymmetrise the conformal anomaly term if one
doubles the number of fermions, by adding an additional doublet
index,
\be
\psi_\mu\longrightarrow \psi_\mu^\alpha\,,\qquad
\lambda\longrightarrow \lambda^\alpha\,.
\ee
All the previous fermion bilinears in the Lagrangian will now have
$\alpha$ and $\beta$ indices contracted with $\delta_{\alpha\beta}$.  The
terms in ${\cal L}_c$, on the other hand, will have the $\alpha$ and $\beta$
indices contracted with $\epsilon_{\alpha\beta}$.  An $\ep_{\alpha\beta}$
should also be inserted in the extra terms (\ref{conftrans}) in
transformation rules for $\psi_\mu$ and $\lambda$.

\section{Conclusions}

We have introduced the notion of a {\it pseudo-supersymmetric}
theory, meaning a theory which is invariant under supersymmetry-like
transformations provided that one works only up to the quadratic
order in fermions.  The symmetry could not be extended beyond the
quadratic level, except in those cases where the theory happens in
fact to be supersymmetric in the usual sense.  A simple example of a
class of pseudosupersymmetric theories is Einstein gravity in an
arbitrary dimension $D$, to which one adds a ``gravitino'' kinetic
term, as in equation (\ref{einstein}).

One can do many of the things with a pseudo-supersymmetric theory
that one normally does with a genuinely supersymmetric one,
including defining bosonic BPS solutions as those for which there
exists one or more Killing spinors for which the fermion variations
vanish.  For example, in the case of pseudo-supersymmetrised
Einstein gravity, the BPS solutions will be Ricci-flat metrics for
which the pseudo-gravitino variation given by (\ref{einsteinsusy})
vanishes.  In other words, the BPS solutions in this case will be
Ricci-flat spaces of special holonomy, where there exists one or
more covariantly-constant spinors.  As well as providing a
characterisation of solutions with additional structures analogous
to special holonomy, imposing the pseudo-supersymmetry requirement
on bosonic backgrounds can also make it easier to construct the
solutions, since it implies that the bosonic fields will be
constrained by systems of first-order differential equations.

In this paper, we have constructed a pseudo-supersymmetric fermionic
extension of the low-energy effective Lagrangian (\ref{boslag}) for
the bosonic string. Only in ten dimensions, it would be possible to
extend the pseudo-supersymmetry to become an exact supersymmetry, by
adding the necessary higher-order fermion terms that would recover
the standard ${\cal N}=1$, $D=10$ supergravity of \cite{Bvn}.  We
then showed that one can couple a Maxwell or Yang-Mills
pseudo-supermultiplet to the pseudo-supergravity in arbitrary
dimensions.  Using an observation made in
\cite{bss,berderoo,berderoo2}, we then showed how the Yang-Mills
result can be used in order to pseudo-supersymmetrise a
curvature-squared $\alpha'$ correction to the theory, giving a
theory that is pseudo-supersymmetric up to the $\alpha'$ order.  We
furthermore showed that dilaton potential term that arises as the
conformal anomaly if the bosonic string is embedded in dimensions
other than 26 can be pseudo-supersymmetrised in an arbitrary
dimension $D$.

Although it is possible to pseudo-supersymmetrise a considerably
wider class of bosonic theories than those that can be genuinely
supersymmetrised, it should be emphasised that it is nonetheless a
highly restrictive requirement on a bosonic theory. It would be very
interesting to try to gain an understanding of what the underlying
properties of a bosonic theory are that allow it to be
pseudo-supersymmetrised. One of the features of the low-energy
effective action of the bosonic string that led us to suspect that
it might admit a pseudo-supersymmetrisation was that the theory
admits remarkable Pauli sphere reductions on $S^3$ or $S^{D-3}$
\cite{clpgen}.  Almost all the other known examples of consistent
Pauli sphere reductions arise in supersymmetric theories. Thus there
does seem to be a possible correlation between pseudo-supersymmetry
and the occurrence of consistent Pauli sphere reductions.  This
could provide an interesting avenue for further investigation.

    The successful construction of the pseudo-supergravity extension
of the bosonic string in this paper suggests that
pseudo-supersymmetry may be an intrinsic stringy property. It is of
great interest to study at the world-sheet level whether one can
construct pseudo-supersymmetric string theory whose low energy
effective action is pseudo-supergravity.  This line of investigation
may yield models that confer many desirable features of
supersymmetry and yet conform with the non-supersymmetric nature of
our world.

\section*{Acknowledgement}

We are grateful to Haishan Liu, Ergin Sezgin and Paul Townsend for
helpful discussions. C.N.P. is grateful to the KITPC, Beijing, for
hospitality during the course of this work. The research of C.N.P.
is supported in part by DOE grant DE-FG03-95ER40917.

\appendix

\section{$\Gamma$ Matrices and Spinors in Dimension $D$}

   Here, we record some basic results about the symmetry properties of
the antisymmetric products $\Gamma^{(n)}\sim \Gamma^{\mu_1\cdots\mu_n}$ of
$\Gamma$ matrices, and about the occurrence of Majorana or
symplectic-Majorana representations, in general dimensions $D$.  We refer the
reader to \cite{vanp} for further details.

  In the table below, we indicate the symmetry properties of $C\Gamma^{(n)}$
for each dimension $D$ mod 8, where $C$ is the charge conjugation matrix.
The symmetry (S=symmetric, A=antisymmetric)
of $C\Gamma^{(n)}$ is always the same as that of
$C\Gamma^{(n+4)}$.  The column headed ``Spinor'' indicates whether the
representation is Majorana (M) or symplectic-Majorana (S-M).

Note that the $\Gamma$-matrix products $C\Gamma^{(n)}$ for odd
values of $n$ always have identical symmetry properties in all
Majorana representations (i.e. $C\Gamma^{(1)}, C\Gamma^{(5)},
\ldots$ are symmetric, and $C\Gamma^{(3)}, C\Gamma^{(7)},\ldots$ are
antisymmetric), and precisely the opposite symmetry properties in
all symplectic-Majorana representations (i.e. $C\Gamma^{(1)},
  C\Gamma^{(5)},\ldots$ are
antisymmetric; $C\Gamma^{(3)}, C\Gamma^{(7)},\ldots$ are
symmetric).  For even values of $n$, on the other
hand, we have to distinguish between two possible symmetry patterns.  They
may be characterised by a parameter $\beta$, which is either $+1$ or $-1$.
It can be defined through the equation
\be
\Gamma_\mu^T = \beta\, C \Gamma_\mu C^{-1}\,.\label{betadef}
\ee
The values of $\beta$ for each representation are listed in Table
1.\footnote{In fact our $\beta=-\eta$, where $\eta=\pm1$ is defined
in \cite{vanp} through $\Gamma_\mu^T = -\eta\, C\, \Gamma_\mu\,
C^{-1}$. We prefer to avoid the symbol $\eta$ because it is often
used to denote a Killing spinor, and also because we find the
opposite sign in our definition of $\beta$ more convenient.}

   In our calculations and results, we assume that all spinors are either
Majorana or symplectic-Majorana, according to the allowed possibilities in
each spacetime dimension $D$.  Symplectic-Majorana spinors each carry a
doublet $Sp(2)$ index $i$, and in any fermion bilinear the two indices are
contacted with the antisymmetric $Sp(2)$-invariant tensor $\epsilon_{ij}$.
By then suppressing the $Sp(2)$ indices, certain fermionic expressions
take the same form in any spacetime dimension, regardless of whether
the fermions
are Majorana or symplectic-Majorana.  For example, the pseudo-gravitino
kinetic term may be written as
\be
\ft12 \bar\psi_\mu \Gamma^{\mu\nu\rho} D_\nu\psi_\rho
\ee
in all cases, where, in the case of symplectic-Majorana spinors, this
means $\ft12 \ep_{ij} \bar\psi^i_\mu \Gamma^{\mu\nu\rho} D_\nu
\psi_\rho^j$.  There are other terms in the pseudo-supersymmetric
extension of the bosonic string whose coefficients depend on the sign
of $\beta$, which can be read off on a case by case basis from the table.

\bigskip\bigskip

\centerline{
\begin{tabular}{|c|c|c|c|c|c|c||c|c|}\hline
 $D$ mod 8 & $C\Gamma^{(0)}$ & $C\Gamma^{(1)}$ & $C\Gamma^{(2)}$ &
$C\Gamma^{(3)}$ & $C\Gamma^{(4)}$ & $C\Gamma^{(5)}$ & Spinor &$\beta$\\
\hline\hline
0 & S & S & A & A & S & S & M & $+1$ \\
  & S & A & A & S & S & A & S-M & $-1$ \\ \hline
1 & S & S & A & A & S & S & M & $+1$ \\ \hline
2 & S & S & A & A & S & S & M & $+1$ \\
  & A & S & S & A & A & S & M & $-1$ \\ \hline
3 & A & S & S & A & A & S & M & $-1$ \\ \hline
4 & A & S & S & A & A & S & M & $-1$ \\
  & A & A & S & S & A & A & S-M & $+1$ \\ \hline
5 & A & A & S & S & A & A & S-M & $+1$ \\ \hline
6 & A & A & S & S & A & A & S-M & $+1$ \\
  & S & A & A & S & S & A & S-M & $-1$ \\ \hline
7 & S & A & A & S & S & A & S-M & $-1$ \\ \hline
\end{tabular}}
\bigskip

\centerline{Table 1: $\Gamma$-matrix symmetries and spinor reperesentations
in diverse dimensions.}

\end{document}